\def\Rfr#1{Eq. (\ref{#1})}

\def\asec{$''$ cy$^{-1}$}

\def\dert#1#2{\frac{{{d}}{#1}}{{{d}}{#2}}}              

\def\asec{$''$ cy$^{-1}$}

\def\bar{\begin{eqnarray}}
\def\ear{\end{eqnarray}}

\def\eqi{\begin{equation}}
\def\eqf{\end{equation}}
\def\eqia{\begin{eqnarray}}
\def\eqfa{\end{eqnarray}}
\def\rp#1#2{{#1\over#2}}

\def\lb#1{\label{#1}}

\def\oc2{$\mathcal{O}(c^{-2})$}

\documentclass{elsart5p}
\usepackage{ifpdf}
\usepackage{graphicx,natbib,amssymb,lineno}
\ifpdf
\usepackage[%
  pdftitle={Instructions for use of the document class
    elsart},%
  pdfauthor={Simon Pepping},%
  pdfsubject={The preprint document class elsart},%
  pdfkeywords={instructions for use, elsart, document class},%
  pdfstartview=FitH,%
  bookmarks=true,%
  bookmarksopen=true,%
  breaklinks=true,%
  colorlinks=true,%
  linkcolor=blue,anchorcolor=blue,%
  citecolor=blue,filecolor=blue,%
  menucolor=blue,pagecolor=blue,%
  urlcolor=blue]{hyperref}
\else
\usepackage[%
  breaklinks=true,%
  colorlinks=true,%
  linkcolor=blue,anchorcolor=blue,%
  citecolor=blue,filecolor=blue,%
  menucolor=blue,pagecolor=blue,%
  urlcolor=blue]{hyperref}
\fi

\makeatletter
\def\elsartstyle{%
    \def\normalsize{\@setfontsize\normalsize\@xiipt{14.5}}
    \def\small{\@setfontsize\small\@xipt{13.6}}
    \let\footnotesize=\small
    \def\large{\@setfontsize\large\@xivpt{18}}
    \def\Large{\@setfontsize\Large\@xviipt{22}}
    \skip\@mpfootins = 18\p@ \@plus 2\p@
    \normalsize
}
\@ifundefined{square}{}{}
\makeatother

\journal{Planetary Space Science}
\begin{document}

\begin{frontmatter}
\title{The impact of the Kuiper Belt Objects and of the asteroid ring on future high-precision relativistic Solar System tests}

\author{Lorenzo Iorio\corauthref{boh}}
\address{Viale Unit$\grave{a}$ di Italia 68, 70125\\Bari (BA), Italy}

\corauth[boh]{Corresponding author}
\ead{lorenzo.iorio@libero.it}

\begin{abstract}
We preliminarily investigate the impact of the Kuiper Belt Objects (KBOs) and of the asteroid ring
on some proposed high-precision tests of Newtonian and post-Newtonian gravity to be performed in the
Solar System   by means of spacecraft in heliocentric $\approx 1$ AU orbits and accurate orbit determination of some of the inner planets. It turns out that the Classical KBOSs (CKBOS), which amount to $\approx 70\%$ of the observed population of Trans-Neptunian bodies, induce a systematic secular error of about 1 m after one year in the transverse direction $T$ of the orbit of a test particle orbiting at 1 AU from the Sun. For Mercury the ratios of the secular perihelion precessions induced by CKBOs to the ones induced by the general relativity and the solar oblateness $J_2$ amount to $6\times 10^{-7}$ and $8\times 10^{-4}$, respectively.
The secular transverse perturbation induced on a $\approx 1$ AU orbit by the asteroid ring, which globally accounts for the action of the minor asteroids whose mass is about $5\times 10^{-10}$ solar masses,  is  $10$ m yr$^{-1}$;
the bias on the relativistic and $J_2$ Mercury perihelion precessions is $6.1\times 10^{-6}$ and $1\times 10^{-2}$, respectively.
Given the very ambitious goals of many expensive and complex missions aimed to testing gravitational theories to  unprecedented levels of accuracy, these notes may suggest further and more accurate investigations of such  sources of potentially insidious systematic bias.
\end{abstract}

\end{frontmatter}

\section{Introduction}
The goal of this letter is to draw attention on the potentially corrupting effect of the Kuiper Belt Objects (KBOs)
and of the ring of the minor asteroids on many proposed high-precision tests of Newtonian and post-Newtonian gravity to be performed in a more or less near future with
spacecraft moving along heliocentric orbits in the inner regions of the Solar System.
\section{The PPN parameters $\beta$ and $\gamma$ and their current experimental bounds}
In the Parameterized Post-Newtonian (PPN) formalism \citep{Wil93},
developed by Nordtvedt and Will to make easier the comparison of
metric theories of gravity with each other and with experiment,
the Eddington-Robertson-Schiff parameters $\beta$ and $\gamma$
describe how much non-linearity is present in the superposition
law for gravity and how much spatial curvature is produced by a
unit mass, respectively; in general relativity $\beta=\gamma=1$.
For a recent overview of the measurements of such PPN parameters
see \citep{Wil06}.

The most accurate and clean determinations of $\gamma$ come from
the effects involving the propagation of electromagnetic signals:
in their PPN expressions only $\gamma$ is present, at first order.
For the Shapiro time delay \citep{Sha64}, recently tested with
the Doppler tracking of the Cassini spacecraft, the most accurate
result is $\gamma-1=(2.1\pm 2.3)\times 10^{-5}$ \citep{Bertotti03}.
With $\beta$ the situation is less favorable because it enters the
post-Newtonian equations of motion of test particles in
conjunction with $\gamma$. For
example, in the post-Newtonian expression of the well known
Einstein precession of the pericentre $\omega$ of a test particle
\citep{Ein15} the combination $\nu=(2+2\gamma-\beta)/3$ is
present. E.V. Pitjeva recently processed more than 317 000 Solar
System observations (1913-2003) for the construction of the
Ephemerides of Planets and the Moon EPM2004 determining, among
other things, $\beta$ and $\gamma$ with an accuracy of $10^{-4}$
\citep{Pit05}. The same limit for $\beta$ can also be obtained by
combining the Cassini result for $\gamma$ with the Lunar Laser
Ranging test of the  Nordtvedt
effect\footnote{It is a violation of the strong equivalence
principle which is zero in general relativity. } \citep{Nor68a, Nor68b}, expressed in terms of $\eta=4\beta-\gamma-3$: indeed, from
$\delta\eta=4.5\times 10^{-4}$ \citep{Bertetal06}, it
follows $\delta\beta=1.3\times 10^{-4}$.
\section{Proposed space-based  missions for unprecedented tests of post-Newtonian gravity}
Various theoretical models involving scalar-tensor scenarios
\citep{Dam93a, Dam93b, Dam96a, Dam96b, Dam02a, Dam02b} predicts deviations
from unity for $\beta$ and $\gamma$ at a $10^{-6}-10^{-7}$ level,
so that it is very important to push the accuracy of such PPN
tests towards this demanding accuracy.
As a consequence, many space-based dedicated missions have been so far proposed; they share the same operational scenario, i.e. the inner regions of the Solar System from 0.38 AU to about 1 AU.
Below we briefly review some of them.

The LATOR (Laser Astrometric Test Of Relativity) mission \citep{Tur06}  has, among its
goals, the determination of $\gamma$ with an accuracy of
$10^{-9}$ from the first order and a direct and independent
measurement of $\beta$ at a $\sim 0.01\%$ level via the
second-order gravity-induced deflection of  light.
LATOR is based on the use of the International Space Station (ISS) and of two small
spacecraft to be placed in heliocentric $\approx 1$ AU 3:2 orbits resonant with that of the Earth.
The nominal mission lifetime amounts to 22 months.

The most accurate method available for determining planetary orbits
is based on accurate range measurements from the Earth to a spacecraft orbiting another planet
or to a lander on the planet's surface. The ESA BepiColombo mission (http://sci.esa.int/science-e/www/area/index.cfm?fareaid=30), which should be launched in 2012,
 places itself in such a framework aiming at measuring $\beta$ and $\gamma$  at a $10^{-6}$ level \citep{Mil02} from an accurate determination
of the orbits of Mercury and the Earth \citep{Ash07}. The official science mission duration is 1 year with possible extensions to 8 years.

The accurate measurement of $\beta$ and $\gamma$ is one
of the scientific goals also of the proposed mission ASTROD (Astrodynamical Space Test of Relativity using Optical Devices) \citep{Ni02, Ni04, Ni06}; the expected accuracy is of the order of $
10^{-7}$ or better for both \citep{Ni04, Ni06}. The concept of ASTROD is to put two spacecraft in separate solar
orbits and carry out laser interferometic ranging with Earth reference stations (e.g. a
spacecraft at the Earth-Sun L1/L2 points). A simple version of ASTROD, ASTROD
I, formerly known as Mini-ASTROD, has been studied as the first step to ASTROD. ASTROD I employs
one spacecraft in a solar orbit and carries out interferometric ranging and pulse
ranging with ground stations. The distance of ASTROD I from the Sun varies from 0.5 AU to 1 AU \citep{Ni04}.
The nominal mission lifetime is about $1-2$ year.

Another very important future mission is LISA (Laser Interferometer Space Antenna)\footnote{See on the WEB http://lisa.esa.int/science-e/www/object/index.cfm?fobjectid=31380}, designed to search for and detect gravitational radiation from astronomical sources. The LISA concept implies the use of three spacecraft orbiting the Sun at 1 AU distance in a quasi-equilateral triangle formation 20 deg behind the Earth due to a compromise between minimising the gravitational disturbances from the Earth-Moon system and the communications needs \citep{Pov06}. The LISA nominal mission is 3.5 years with an extended duration of 8.5 years.
\section{The impact of the Kuiper Belt Objects and of the minor asteroids}
In order to reach their very ambitious goals, the heliocentric orbits of the spacecraft of such missions, and of the Earth itself, must be known with high accuracy; any mismodelled or unmodelelled force acting on them may represent a potential systematic bias to be carefully accounted for. To date, a dynamical feature which is not yet included in the force models of the data reduction softwares used for orbit determination purposes is represented by the KBOs. Attempts towards the implementation of this goal started recently  \citep{Pit06, Sta06}.
\citet{Ior07} explicitly worked out the secular effects induced by the KBOs on the perihelia of the inner planets pointing out that they may affect accurate tests of Einsteinian and post-Einsteinian gravity to be performed in that regions of the Solar System. Following the approach outlined in \citep{Ber06, Ior07}, it is possible to analytically calculate the cumulative shift in the transverse orbital direction $T$ over a time span $\Delta P$. In the planet's small eccentricity and inclination limit, the Classical KBOs (CKBOs), which follow nearly circular orbits with relatively low eccentricities ($e\leq 0.25$) and semimajor axes $41\lesssim a\lesssim 46$ AU, constituting about $70\%$ of the so far observed population,  induce a transverse perturbation
\eqi \Delta T\approx d\Delta\lambda=-\rp{\Delta P}{2}\sqrt{\rp{Gd^5}{M_{\odot}}}\rp{m_{\rm K}(2+3\varepsilon^2)}{(R_{\rm max}+R_{\rm min})R_{\rm max}R_{\rm min}},\lb{rtu}\eqf where $\lambda$ is the planetary mean longitude,
$G$ is the Newtonian gravitational constant, $M_{\odot}$ is the Sun's mass, $d$ and $\varepsilon$ are the planet's semimajor axis and eccentricity, respectively, $m_{\rm K}$ is the CKBOs' mass, $R_{\rm max}$ and $R_{\rm min}$ are the outer and inner radius, respectively, of the uniform disk with which CKBOs have been modelled \citep{Ber06}. \Rfr{rtu} comes from the integration over one orbital period of the perturbative Gauss equations for the variation of the Keplerian elements in terms of which $\lambda$ can be expressed \citep{Mil87}; since for almost circular and equatorial orbits $\lambda\approx \mathcal{M}+\Omega+\omega$, where $\mathcal{M}$ is the mean anomaly and $\Omega$ is the node, we used \eqi \dert\lambda t\approx n-\rp{2A_r r}{nd^2},\eqf where $n=\sqrt{GM_{\odot}/d^3}$ is the Keplerian mean motion and $A_r$ is the radial perturbing acceleration
\citep{Ber06, Ior07}.

\Rfr{rtu} and $m_{\rm K} = 0.052 m_{\oplus}$ \citep{Ior07} tell us that the nominal values of the transverse shift due to CKBOs amounts to -0.10 m for Mercury and -1.01 m for the Earth after one year; this corresponds to an average CKBOs acceleration of the order of $10^{-15}$ m s$^{-2}$ at $r\approx 1$ AU. Since the uncertainty in the CKBOs' mass is high, $\delta m_{\rm K} = 0.223 m_{\oplus}$  (Iorio, 2007) yields upper bounds of 0.42 m for Mercury and 4.33 m for the Earth after one year. In fact, a uniform disk like the one with which the action of the CKBOs is modelled induces not only a radial acceleration \citep{And02, Nie05, Ber06}  but also a normal one \citep{Owe03}. Thus, also the out-of-plane portion $N$ of the orbit
of a planet/spacecraft is, in principle, affected increasing the total orbit bias. However, a quantitative evaluation of such a further bias is beyond the scope of the present letter.

Another way to give an order-of-magnitude assessment of the bias due to CKBOs on the recovery of the relativistic effects of interest, especially useful for BepiColombo since it will use, among other things, the orbital motions of Mercury and the Earth, is to consider the ratio of the perihelion rates induced by CKBOs \citep{Ior07} to the general relativistic ones: it amounts to $6\times 10^{-7}$ for Mercury and $3\times 10^{-5}$ for the Earth. It may also be interesting to do the same for the precessions induced by the solar quadrupole mass moment $J_2$, whose measurement at $10^{-9}$ level of accuracy, or better, is one of the goals of BepiColombo \citep{Mil02, Ash07}: according to \citet{Ior07} and by assuming $J_2=2\times 10^{-7}$, the bias due to the CKBOs is $8\times 10^{-4}$ and $1\times 10^{-1}$ for Mercury and the Earth, respectively.
In Table \ref{TAV} we release the details.

The same reasonings can be made for the hidden mass in the asteroid belt \citep{Kra02} whose action does not include the individual perturbations of the 342 major  asteroids and can be modelled as a massive ring with a constant mass distribution in the ecliptic plane. By assuming\footnote{\citet{Kon06} obtained a smaller mass for the asteroid ring.} \citep{Kra02} $m_{\rm ring}=(5\pm 1)\times 10^{-10}M_{\odot}$ and $R\approx 2.80$ AU, with a $3\%$ uncertainty, we have
\eqi\Delta T^{(\rm ring)} = 10.7\pm 1\ {\rm m}\eqf for a $\approx 1$ AU orbit after one year. The ratios of the perihelion rates induced by the asteroid ring to the general relativistic and $J_2$ ones amount to $(6.1\pm 1.2)\times 10^{-6}$ and $(1\pm 0.2)\times 10^{-2}$ for Mercury, and $(3\pm 0.6)\times 10^{-4}$ and $1.25\pm 0.25$ for the Earth, as can be noted from Table \ref{TAV}.

Given the very ambitious goals of the missions examined, the figures obtained here may yield some further reasons to perform extensive numerical simulations and to enforce the process of accurate modelling of the action of KBOs and of the minor asteroids in planetary/spacecraft orbit determination softwares. In particular, it would be useful to estimate the KBOs mass along with the rest of parameters in a global, least-square solution of the Solar System dynamics using a massive data set to investigate, among other things, the correlation between the parameters of the KBO and of the asteroid belt.

\section*{Acknowledgments}
I gratefully thank S. G. Turyshev, NASA Jet Propulsion Laboratory
(JPL), and A. Povoleri, Astrium EADS, for interesting discussions. Thanks also to an anonymous referee whose comments improved this manuscript.


%
%
%
%
%
\newpage
\begin{table}
\caption{ Nominal values of the secular perihelion precessions $\dot\omega$ of Mercury and the Earth, in arcseconds per century (\asec), induced by general relativity (GR), solar quadrupole mass moment ($J_2$), CKBOs and the asteroid ring.  }\label{TAV}

\begin{tabular}{lllll} \noalign{\hrule height 1.5pt}
Planet & GR & $J_2$ & CKBOs & asteroid ring\\
\hline
Mercury & 42.98047 & 0.02542 &  0.00002 & 0.00026\\
Earth & 3.83869 & 0.00084 &  0.00010 & 0.00110\\

\hline

\noalign{\hrule height 1.5pt}
\end{tabular}

\end{table}

\end{document}